\documentclass[11pt,a4paper]{article}
\usepackage{amssymb}
\usepackage{amsmath}
\usepackage{bm}
\usepackage{color}
\RequirePackage[numbers,sort&compress]{natbib}

\def\onehalf{\textstyle{\frac{1}{2}}}

\def\daw{{\stackrel{\bullet}{\Box}}{}}
\def\Aw{{\stackrel{~\bullet}{A}}{}}

\def\nablaw{{\stackrel{\bullet}{\nabla}}{}}

\def\be{\begin{equation}}
\def\ee{\end{equation}}
\def\ba{\begin{eqnarray}}
\def\ea{\end{eqnarray}}

\usepackage[colorlinks=true, pdfstartview=FitV, linkcolor=blue, 
            citecolor=blue, urlcolor=blue]{hyperref}
\begin{document}

\renewcommand{\thefootnote}{\fnsymbol{footnote}}
\begin{center}
{\Large \bf Conformal and gauge invariant spin-2 field equations}
\vskip 0.5cm
{\bf C. S. O. Mayor, G. Otalora and J. G. Pereira}
\vskip 0.03cm 
{\it Instituto de F\'{\i}sica Te\'orica, UNESP-Univ Estadual Paulista \\
Caixa Postal 70532-2, 01156-970 S\~ao Paulo, Brazil}
\end{center}
\vskip 0.5cm
\begin{quote}
{\footnotesize Using an approach based on the Casimir operators of the de Sitter group, the conformal invariant equations for a fundamental spin-2 field are obtained, and their consistency discussed. It is shown that, only when the spin-2 field is interpreted as a 1-form assuming values in the Lie algebra of the translation group, rather than a symmetric second-rank tensor, the field equation is both conformal and gauge invariant.

}
\end{quote}



\section{Introduction}
\label{intro}

Considering that the light-cone is conformal invariant, the equation governing the dynamics of a massless field --- which lives in the light-cone --- must also be conformal invariant \cite{wald}. A trivial example is the electromagnetic field, whose field equations are well-known to be conformal invariant; the massless Dirac equation is another relevant example. The conformal invariant equations for all other massless fields can only be obtained through the introduction of non-minimal couplings of the field to the spacetime curvature. This holds in particular for a symmetric second-rank spin-2 field \cite{CInv2,CInv3,CInv4}. Such field, however, shows consistency problems when coupled to gravitation \cite{AraDeser}. One of these problems is that the gravitationally coupled field equation is no longer gauge invariant, and the spurious degrees of freedom cannot be removed.

On the other hand, a symmetric second-rank tensor is not the only way to represent a spin-2 field. In fact, it can also be represented by a 1-form assuming values in the translation group \cite{spin2}. This doubly representation is related to the fact that gravitation has, in addition to the usual metric formulation, also a tetrad formulation. Accordingly, whereas a symmetric second-rank field would be a field conceptually similar to a metric perturbation, a 1-form assuming values in the translation group would be conceptually similar to a perturbation of the tetrad. Of course, both variables represent the very same spin-2 field.

With these points in mind, the purpose of this paper is to make an analysis of the field equations representing a fundamental massless, traceless spin-2 field. First, we are going to obtain the conformal invariant equations for both a symmetric second-rank field and for a 1-form assuming values in the translation group. This will be done by making use of a new method of obtaining conformal invariant equations, which is based on the Casimir operator of the de Sitter group. Then, a study of the gauge and conformal invariance of these equations will be performed. The basic conclusion is that, only when the spin-2 field is interpreted as a 1-form assuming values in the Lie algebra of the translation group, the ensuing field equation is both conformal and gauge invariant. This result suggests that, instead of a symmetric second-rank tensor, a fundamental spin-2 field should be represented by a translational-valued 1-form, a result that is in consonnance with the principles of teleparallel gravity \cite{spin2}.

\section{Field equations in Minkowski spacetime}
\label{FEMink}

In Minkowski spacetime, the field equations for any field belonging to an irreducible representation of the Poincar\'e group can be obtained from the first Casimir invariant of the Poincar\'e group,\footnote{The Greek alphabet $\mu, \nu, \rho, \dots = 0,1,2,3$ will be used to denote both spacetime and algebraic indices related to the de Sitter group.}
\be
\hat{\mathcal C}_{P} = \eta^{\mu \nu} \hat{P}_\mu \hat{P}_\nu,
\label{FirstCas}
\ee
where $\hat{P}_\mu = \partial_\mu$. Considering that the eigenvalues ${\mathcal C}_{P}$ of the Casimir operator $\hat{\mathcal C}_{P}$ are given by
\be
{\mathcal C}_{P} = - m^2,
\label{autovacas}
\ee
with $m$ the mass of the field, we see that a general tensorial field $\Psi_{\mu_1 ... \mu_{\sf s}}$ of spin ${\sf s}$ satisfies the equation
\be
\hat{\mathcal C}_{P} \Psi_{\mu_1 ... \mu_{\sf s}} = - m^2 \Psi_{\mu_1 ... \mu_{\sf s}}.
\ee
From now on our interest will be concentrated only on massless, traceless fields. Accordingly, instead of considering the spin ${\sf s}$, we are going to consider the modulus of the helicity $\sigma = |{\bm \sigma}|$ of the fields. Recall that the spin of a given representation $(A, B)$ of the Lorentz group is defined by
\[
{\sf s} = A+B,
\]
whereas the helicity is \cite{weinberg2}
\[
{\bm \sigma} = \pm |B - A|.
\]

A crucial point of the Casimir approach is that it gives rise to field equations already in the Lorenz gauge. In order to obtain the equations without any gauge choice, it is necessary to re-introduce the Lorenz gauge into the field equations. This can be done by adding to them a gauge term of the form $\partial_{\mu_{\sigma}} \partial^\rho \Psi_{\mu_1 ... \mu_{\sigma} \rho}$, in such a way that they read
\be
\hat{\mathcal C}_{P} \Psi_{\mu_1 ... \mu_\sigma} + 
a \, \partial_{\mu_\sigma} \partial^\rho \Psi_{\mu_1 ... \mu_{\sigma-1} \rho} = 0,
\ee
where \cite{EO}
\be
a = - \frac{2 \sigma}{\sigma + 1}
\label{a}
\ee
is a constant that depends on the helicity $\sigma$ of the field. This is the field equation satisfied by a general bosonic field of helicity $\sigma$ in Minkowski spacetime.

\section{The de Sitter spacetime and group}
\label{deSitterSG}

\subsection{de Sitter spacetime}

The de Sitter spacetime can be defined as a hyper-surface in the ``host'' pseudo-Euclidean space ${\bf E}^{4,1}$, inclusion whose points in Cartesian coordinates $\chi^A$ $(A, B, ... = 0, ..., 4)$ satisfy \cite{ellis}
\[
\eta_{AB} \chi^A \chi^B = - \, l^2,
\]
where $\eta_{AB}$ = $(1,-1,-1,-1,-1)$ and $l$ is the de Sitter length-parameter (or pseudo-radius). It has the pseudo-orthogonal group $SO(4,1)$ as group of motions. The four-dimensional stereographic coordinates $x^\mu$ are obtained by performing a stereographic projection from the de Sitter hyper-surface 
\be
\eta_{\mu \nu} \, \chi^{\mu} \chi^{\nu} - (\chi^{4})^2 = -\, l^2
\label{dspace1}
\ee
into a target Minkowski spacetime. They are defined by~\cite{gursey}
\be
\chi^{\mu} = \Omega \, x^\mu \quad \mbox{and} \quad \chi^4 = -\, l \, \Omega \left(1 + 
{\zeta^2}/{4 l^2} \right),
\label{stepro}
\ee 
where
\be
\Omega = \frac{1}{1 - {\zeta^2}/{4 l^2}}, 
\label{n}
\ee
with $\zeta^2$ the Lorentz-invariant quadratic interval $\zeta^2 = \eta_{\mu \nu} \, x^\mu x^\nu$. In these coordinates, the de Sitter line element $ds^2 = \eta_{AB} \, d\chi^A d\chi^B$ reduces to $ds^2 = g_{\mu \nu} \,d x^\mu dx^\nu$, with
\be
g_{\mu \nu} = \Omega^2 \, \eta_{\mu \nu}
\label{44}
\ee
the de Sitter metric. The de Sitter spacetime, therefore, is conformally flat, with the conformal factor given by $\Omega^2$. The Christoffel connection of the de Sitter metric (\ref{44}) is \cite{livro}
\be
\Gamma^{\lambda}{}_{\mu \nu} = \left[ \delta^{\lambda}{}_{\mu}
\delta^{\rho}{}_{\nu}  + \delta^{\lambda}{}_{\nu}
\delta^{\rho}{}_{\mu} - \eta_{\mu \nu} \eta^{\lambda \rho} \right]
\partial_\rho \left(\ln \Omega \right).
\label{46}
\ee
The corresponding Riemann tensor components, are found to be
\be
R^{\mu}{}_{\nu \rho \sigma} = - \, \frac{1}{l^2} \,
\left(\delta^{\mu}{}_{\rho} g_{\nu \sigma} - \delta^{\mu}{}_{\sigma} g_{\nu
\rho} \right).
\label{47}
\ee
The Ricci tensor and the scalar curvature are, respectively,
\be
R_{\nu \sigma} =  - \, \frac{3}{l^2} \, g_{\nu \sigma} \quad \mbox{and} \quad
R =  -\, \frac{12}{l^2}.
\label{49}
\ee

\subsection{de Sitter transformations}

In terms of the host pseudo-Euclidian coordinates $\chi^A$, an infinitesimal de Sitter transformation is defined by
\be
\delta \chi^C = \onehalf \, {\mathcal E}^{AB} \hat{L}_{AB} \, \chi^C,
\label{dSHost}
\ee
where ${\mathcal E}^{AB} = - {\mathcal E}^{BA}$ are the transformation parameters, and
\be
\hat{L}_{AB} = \eta_{AC} \, \chi^{C} \frac{\partial}{\partial \chi^{B}} -
\eta_{BC} \, \chi^{C} \frac{\partial}{\partial \chi^{A}}
\ee
are the de Sitter generators. In terms of the stereographic coordinates, the infinitesimal de Sitter transformations (\ref{dSHost}) assume the form
\be\label{ds-t}
\delta x^\mu = \onehalf \epsilon^{\rho \sigma} \hat{L}_{\rho \sigma} x^\mu -
\epsilon^\rho \hat{\Pi}_\rho x^\mu,
\ee
where
$\epsilon^{\rho \sigma} = {\mathcal E}^{\rho \sigma}$ {and} $\epsilon^{\rho} = l \, {\mathcal E}^{\rho 4}$ are the transformation parameters,
\be\label{lorentzdS}
\hat{L}_{\rho \sigma} =
\eta_{\rho \lambda} \, x^\lambda \, \hat{P}_\sigma -
\eta_{\sigma \lambda} \, x^\lambda \, \hat{P}_\rho
\ee
are the Lorentz generators, and
\be\label{pi}
\hat{\Pi}_\rho \equiv \frac{\hat L_{\rho 4}}{l} =
\hat P_\rho - \frac{1}{4 l^2} \, \hat K_\rho
\ee
are the so-called de Sitter ``translation'' generators, with
\be
\hat{P}_\rho = \partial/ \partial x^\rho
\label{TransGenerators1}
\ee
the translation generators, and
\be
\hat{K}_\rho = \left(2 \eta_{\rho \nu} \, x^\nu x^\mu - \zeta^2 \delta_{\rho}{}^{\mu} \right) 
\partial/ \partial x^\mu,
\label{TransGenerators2}
\ee
the generators of proper conformal transformations \cite{coleman}. 

\section{Conformal invariant field equations}
\label{FEdS}

\subsection{The Casimir operator approach}

The de Sitter spacetime is transitive under a combination of translations and proper conformal transformations \cite{dStransi}. If instead of Minkowski the spacetime representing absence of gravitation is de Sitter, the group governing the spacetime kinematics must change from Poincar\'e to de Sitter group. Since the latter includes the proper conformal transformations, the field equations obtained from the first Casimir operator of the de Sitter group are naturally conformal invariant. Remember that under the conformal re-scaling of the metric\footnote{In the specific case of a conformal transformation leading the Minkowski metric $\eta_{\mu \nu}$ to the de Sitter metric~(\ref{44}), the conformal factor $\Omega$ is that given by Eq.~(\ref{n}).}
\be
\bar{g}_{\mu \nu} = \Omega^{2} \, g_{\mu \nu},
\label{CoMe}
\ee
where $\Omega = \Omega(x)$ is the conformal factor, a general bosonic field transforms according to \cite{bd} 
\be
\bar{\Psi} = \Omega^{\sigma - 1} \, \Psi.
\label{CoField}
\ee
The power of $\Omega$ is known as the conformal weight of the field $\Psi$.

Let us then obtain the conformal invariant field equation for a general massless bosonic field $\Psi$ with helicity $\sigma$. The first Casimir invariant of the de Sitter group is
\be
\hat{\mathcal C}_{dS} = -\frac{1}{2 l^2} \, \eta^{AC} \, \eta^{BD} \, 
\hat{J}_{AB} \, \hat{J}_{CD},
\label{1dsCas}
\ee
where $\hat{J}_{AB}$ are the de Sitter generators. In stereographic coordinates, it can be written as
\be
\hat{\mathcal C}_{dS} =  -\frac{1}{2l^2} \left( \eta^{\alpha \beta} \, \eta^{\gamma \delta} \, \hat J_{\alpha \gamma} \, \hat J_{\beta \delta} -
2 \eta^{\alpha \beta} \, \hat{J}_{4 \alpha} \hat{J}_{4 \beta}\right).
\label{2dsCas}
\ee
These generators can be decomposed into orbital and spin parts
\be
\hat J_{\alpha \gamma} = \hat L_{\alpha \gamma} + \hat S_{\alpha \gamma}
\qquad \mbox{and} \qquad
\hat J_{4 \gamma} = l ( \hat \Pi_{\gamma} + \hat{\Sigma}_{\gamma} ).
\label{l+s}
\ee
The orbital generators $\hat{L}$ and $\hat{\Pi}$, given respectively by Eqs.~(\ref{lorentzdS}) and (\ref{pi}), are the same for all fields. The explicit form of the matrix generators $\hat S$ and $\hat{\Sigma}$, on the other hand, depend on the spin (or helicity) of the field.
In terms of these generators, the Casimir operator (\ref{2dsCas}) assumes the form
\be
\hat{\mathcal C}_{dS} = - \frac{1}{2l^2} \left( \hat L^2 + \hat L \cdot \hat S +
\hat S \cdot \hat L + \hat S^2 \right) + \hat{\Pi}^2 +
\hat{\Pi} \cdot \hat{\Sigma} + \hat{\Sigma} \cdot \hat{\Pi} + \hat{\Sigma}^2.
\label{GenCasOp}
\ee
For particles belonging to representations on the {\it principal series}, the eigenvalues ${\mathcal C}_{dS}$ of the Casimir operator $\hat{\mathcal C}_{dS}$ are, in the massless case, given in terms of the helicity $\sigma$ by \cite{dix}
\begin{equation}
{\mathcal C}_{dS} = \frac{1}{l^2} \left[\sigma(\sigma+1)-2 \right].
\label{rforcs}
\end{equation}
From the identity $\hat{\mathcal C}_{dS} \Psi_{\mu_1 ... \mu_\sigma} = {\mathcal C}_{dS} \Psi_{\mu_1 ... \mu_\sigma}$, the massless field equation is then found to be
\be
\hat{\mathcal C}_{dS} \Psi_{\mu_1 ... \mu_\sigma}
= - \frac{R}{12} \, [\sigma(\sigma+1)-2] \Psi_{\mu_1 ... \mu_\sigma},
\label{ConInvFE}
\ee
with $R$ the scalar curvature of the de Sitter spacetime. 

Now, similarly to the Minkowski case, the Casimir operator of the de Sitter spacetime gives rise to field equations already in the Lorenz gauge. Since the Lorenz gauge is not conformal invariant, the field equation (\ref{ConInvFE}) is not conformal invariant either. As discussed in Section~2, it is then necessary to re-introduce the Lorenz gauge into the field equation,
\be
\hat{\mathcal C}_{dS} \Psi_{\mu_1 ... \mu_\sigma} + 
a \, \nabla_{\mu_\sigma} \nabla^\rho \Psi_{\mu_1 ... \mu_{\sigma-1} \rho} = 
 - \frac{R}{12} \, [\sigma(\sigma+1)-2] \Psi_{\mu_1 ... \mu_{\sigma}},
\label{ConInvFE2}
\ee
with $a$ given by Eq.~(\ref{a}), and $\nabla_\rho$ the covariant derivative in the Christoffel connection of the de Sitter spacetime. This is the conformal invariant field equation in de Sitter spacetime. Considering that it is covariant, if it is true in de Sitter spacetime, it will be true in any pseudo-Riemannian spacetime with non-constant curvature.

\subsection{The scalar field as an example}

As an illustration of the approach, we are going to obtain the well-known conformal invariant equation for the scalar field $\phi$, for which $\hat{S} \phi = 0$ and $\hat{\Sigma} \phi = 0$. In this case, the Casimir operator (\ref{GenCasOp}) reduces to
\begin{equation}
\hat{\mathcal C}_{dS} \equiv - \frac{1}{2l^2} \hat L^2 + \hat \Pi^2 = 
\Omega^{-2} \eta^{\alpha \beta} \partial_\alpha \partial_\beta +
\frac{1}{l^2} \, \Omega^{-1} x^\alpha \partial_\alpha,
\label{casimirscal}
\end{equation}
with $\Omega$ the conformal factor (\ref{n}). As a simple inspection shows, it can be rewritten in the form
\be
\hat{\mathcal C}_{dS} = g^{\alpha \beta} \nabla_\alpha \nabla_\beta \equiv \Box,
\label{casimirscal2}
\end{equation}
with $\nabla_\alpha$ the covariant derivative in the Christoffel connection of the de Sitter metric, and $\Box$ the corresponding Laplace-Beltrami operator for a scalar field. For $\sigma = 0$, therefore, the field equation (\ref{ConInvFE2}) for $\phi$ is found to be
\begin{equation}
\Box \phi - \frac{R}{6} \, \phi = 0,
\label{cskg}
\end{equation}
with $R$ the scalar curvature of the de Sitter spacetime. Although obtained in de Sitter spacetime, since it is invariant under general coordinate transformations, it will be true in any pseudo-Riemannian spacetime with non-constant curvature. In fact, as is well known, it represents the conformal invariant equation for a scalar field~\cite{bd}.

\section{Spin-2 conformal invariant field equation}
\label{ConfInvS2}

\subsection{The spin-2 Casimir operator}

The dynamics of a fundamental spin-2 field in Minkowski spacetime is expected to coincide with the dynamics of a linear perturbation of the metric $\psi_{\mu \nu}$ around flat spacetime:
\be
g_{\mu \nu} = \eta_{\mu \nu} + \psi_{\mu \nu}.
\ee
For this reason, a fundamental spin-2 field is usually assumed to be described by a symmetric, second-rank tensor $\psi_{\mu \nu} = \psi_{\nu \mu}$. Let us then obtain the Casimir operator for a symmetric, second-rank spin-2 field $\psi_{\mu \nu}$. It is written as
\be
(\hat{C}_{ds})^{\mu\nu}_{\alpha\beta} = - \frac{1}{2l^2} \eta^{AB} \eta^{CD}
(\hat{J}_{AC})^{\mu\nu}_{\rho\sigma} \, (J_{BD})^{\rho\sigma}_{\alpha\beta},
\label{Cas1}
\ee
where generators with {\em four indices} are the spin-2 generators. These generators can be decomposed into orbital and spin parts,
\be
(\hat{J}_{AB})^{\mu\nu}_{\rho\sigma} = (\hat{L}_{AB})^{\mu\nu}_{\rho\sigma} + 
(\hat{S}_{AB})^{\mu\nu}_{\rho\sigma}.
\ee
The orbital generator is the same for all fields, and is given by
\be
(\hat{L}_{AB})^{\mu \nu}_{\rho \sigma} = (\hat{L}_{AB}) \delta^\mu_\rho \, \delta^\nu_\sigma .
\ee
The spin generators, on the other hand, are obtained by summing two spin-1 representations, one for each index of $\psi_{\mu \nu}$,
\be
(\hat{S}_{AB})^{\mu \nu}_{\rho \sigma} = (\hat{S}_{AB})^\mu_\rho \, \delta^\nu_\sigma + 
(\hat{S}_{AB})^\nu_\sigma \, \delta^\mu_\rho,
\label{spin1Gene}
\ee
where the generators with {\em two indices} are the spin-1 generators. In terms of these generators, the Casimir operator (\ref{Cas1}) assumes the form
\ba
(\hat{C}_{ds})^{\mu\nu}_{\alpha\beta} = - \frac{1}{2l^2} \eta^{AB} \eta^{CD} 
\Big[ (\hat L_{AC}) (\hat L_{BD}) \delta^\mu_\alpha \delta^\nu_\beta + 
(\hat L_{AC}) (\hat S_{BD})^\mu_\alpha \, \delta^\nu_\beta +
(\hat L_{AC}) (\hat S_{BD})^\nu_\beta \, \delta^\mu_\alpha  \nonumber \\
+ 2 (\hat S_{AC})^\mu_\alpha (\hat L_{BD}) \delta^\nu_\beta +
2 (\hat S_{AC})^\nu_\beta (\hat L_{BD}) \delta^\mu_\alpha +
(\hat S_{AC})^\nu_\gamma (\hat S_{BD})^\gamma_\beta \, \delta^\mu_\alpha  \nonumber \\
+ (\hat S_{AC})^\mu_\gamma (\hat S_{BD})^\gamma_\alpha \, \delta^\nu_\beta + 
2 (\hat S_{AC})^\mu_\alpha (\hat S_{BD})^\nu_\beta \Big].
\label{CasSpin2}
\ea
All terms in the right-hand side that involves an orbital generator $\hat L_{AC}$ will contribute to the Laplace-Beltrami operator. On the other hand, terms involving two spin generators $\hat S_{AC}$ will contribute with a non-minimal coupling between the field and the spacetime curvature. More specifically, those bearing one Kronecker delta will contribute with terms involving the Ricci tensor. Since the last term does not involve any Kronecker delta, it can only contribute with a term involving the scalar curvature. 

\subsection{Matrix representation of the conformal transformation}

Before proceeding further, it is necessary to obtain the matrix representation of the proper con\-formal transformation. This can be done from the vanishing of the Lie derivative along the de Sitter Killing vectors,
\begin{equation}
(\mathcal{L}_X g)_{\alpha \beta} = X^\gamma \partial_\gamma  g_{\alpha \beta} +
\partial_\alpha  X^\gamma  g_{\gamma \beta} +
\partial_\beta  X^\gamma  g_{\gamma \alpha} = 0,
\end{equation}
where $g_{\alpha \beta}$ is the de Sitter metric. In this expression, $X^\gamma$ is either the Lorentz Killing vector
\be
L^\gamma \equiv \epsilon^{\alpha \beta} L^\gamma_{\alpha \beta} = 
\epsilon^{\alpha \beta} \Big(\eta_{\alpha \delta} x^\delta \delta^\gamma_\beta -
\eta_{\beta \delta} x^\delta \delta^\gamma_\alpha \Big),
\ee
or the de Sitter ``translation'' Killing vector
\be
\Pi^\gamma \equiv \epsilon^\alpha  \Pi^\gamma_\alpha =
\epsilon^\alpha \Big[\delta^\gamma_\alpha -
\frac{1}{4l^2}(2 \eta_{\alpha \delta} x^\delta  x^\gamma -
\zeta^2 \delta^\gamma_\alpha)\Big],
\ee
with $\epsilon^{\alpha \beta} = - \epsilon^{\beta\alpha}$ and $\epsilon^\alpha$ the ten parameters of the de Sitter group. They satisfy the algebra
\begin{eqnarray}
\left[\hat L_{\alpha \beta}, \hat L_{\gamma \delta}\right] &=& 
\eta_{\beta\gamma} \hat L_{\alpha\delta} + \eta_{\alpha\delta} \hat L_{\beta\gamma} -
\eta_{\beta\delta} \hat L_{\alpha\gamma} - \eta_{\alpha\gamma} \hat L_{\beta\delta} \\
\left[\hat{\Pi}_{\alpha}, \hat L_{\gamma \delta}\right] &=& \eta_{\alpha \gamma} \,
\hat \Pi_{\delta} - \eta_{\alpha \delta} \, \hat \Pi_{\gamma} \\
\left[\hat \Pi_{\alpha}, \hat \Pi_{\gamma}\right] &=& \frac{1}{l^2} \, \hat L_{\alpha \gamma},
\end{eqnarray}
where
\be
\hat L_{\alpha \beta} \equiv L^\gamma_{\alpha \beta} \partial_\gamma =
\eta_{\alpha\delta} x^\delta \partial_\beta - \eta_{\beta\delta} x^\delta \partial_\alpha
\label{lab}
\ee
and
\be
\hat \Pi_{\alpha} \equiv \Pi^\gamma_\alpha \partial_\gamma =
\partial_\alpha - \frac{1}{4l^2} (2 \eta_{\alpha \delta} x^\delta  x^\gamma -
\zeta^2 \delta^\gamma_\alpha) \partial_\gamma.
\label{pia}
\ee

In order to obtain the matrix representations appropriate for a vector field $\psi_\mu$, it is necessary to compute the Lie derivative of the field along the direction of the de Sitter Killing vectors. These derivatives are given by
\be
\delta_L \psi_\mu \equiv (\mathcal{L}_L \psi)_\mu =
\epsilon^{\alpha\beta} L^\gamma_{\alpha\beta} \partial_\gamma \psi_\mu +
\epsilon^{\alpha\beta} \partial_\mu L^\gamma_{\alpha\beta} \psi_\gamma
\ee
and
\be
\delta_\Pi \psi_\mu \equiv (\mathcal{L}_\Pi \psi)_\mu =
\epsilon^\alpha \Pi^\gamma_{\alpha} \partial_\gamma \psi_\mu +
\epsilon^\alpha \partial_{\mu} \Pi^\gamma_{\alpha} \psi_\gamma.
\ee
The first term on the right-hand side of these two equations represent the action of the orbital generators. The last term, on the other hand, represent the action of the spin matrix generators. From these terms, therefore, we get the matrix representations
\be
(\hat{S}_{\alpha\beta})_\mu{}^\gamma \psi_\gamma \equiv
\partial_\mu L^\gamma_{\alpha\beta} \psi_\gamma =
(\eta_{\alpha\mu} \delta_\beta{}^\gamma - \eta_{\beta\mu} \delta_\alpha{}^\gamma) \psi_\gamma ~
\label{dSM1}
\ee
and
\be
(\hat{\Sigma}_{\alpha})_\mu{}^\gamma \psi_\gamma
\equiv \partial_{\mu} \Pi^\gamma_{\alpha} \psi_\gamma =
\frac{1}{2l^2}( \eta_{\mu\beta} x^\beta \delta_\alpha{}^\gamma -
\eta_{\alpha\mu} x^\gamma - \eta_{\alpha \beta} x^\beta \delta_\mu{}^\gamma) \psi_\gamma.
\label{dSM2}
\ee
Equation (\ref{dSM1}) is the usual spin-1 matrix representation of the Lorentz group. Equation (\ref{dSM2}) is the required matrix representation of the proper conformal transformations. It is interesting to remark that, whereas ordinary translations do not have a matrix representation, the proper conformal transformations do have it.

\subsection{Conformal invariant field equation}

Using the representations (\ref{lab}, \ref{pia}, \ref{dSM1}, \ref{dSM2}), through a lengthy but straightforward calculation, Eq.~(\ref{CasSpin2}) reduces to
\be 
(\hat{C}_{ds})^{\alpha\beta}_{\mu\nu} = 
\Box \, \delta^\alpha_{(\mu} \delta^\beta_{\nu)} - 
2 \, R^\alpha{}_{(\mu} \delta^\beta_{\nu)} -
\frac{1}{6} \, R \, \delta^\alpha_{(\mu} \, \delta^\beta_{\nu)} +
\frac{2}{3} \, R_{\mu\nu} \, g^{\alpha\beta},
\label{CasSpin2bis}
\ee
with the parentheses indicating symmetrization with a factor $1/2$. Now, the symmetric second-rank potential $\psi_{\mu \nu}$ describes waves with helicity $\sigma = \pm 2$. Its conformal transformation is consequently
\be
\bar{\psi}_{\mu \nu} = \Omega \, {\psi}_{\mu \nu}.
\ee
For a traceless field, $\psi \equiv \psi^\alpha{}_\alpha = 0$, it is an easy task to verify that the field equation~(\ref{ConInvFE2}) with $\sigma = 2$ yields in this case
\be
\Box \, \psi_{\mu \nu} - 
\frac{4}{3} \, \nabla_{(\mu} \nabla^\alpha \psi_{|\alpha| \nu)} - 
2 \, R^\alpha{}_{(\mu} \, \psi_{|\alpha| {\nu)}} + 
\frac{1}{6} \, R \, \psi_{\mu \nu} = 0,
\ee
where the vertical bars indicate that the enclosed index is not included in the symmetrization.
Using the identity
\be
[\nabla_{(\mu}, \nabla^\alpha] \psi_{|\alpha| \nu)} = -
R^\alpha{}_{(\mu} \psi_{|\alpha| \nu)} + 
R^\alpha{}_{(\mu}{}^\gamma{}_{\nu)} \, \psi_{\alpha \gamma},
\ee
it can be rewritten in the form
\be
\Box \, \psi_{\mu \nu} - 
\frac{4}{3} \, \nabla^\alpha \nabla_{(\mu} \psi_{|\alpha| \nu)} - 
\frac{2}{3} \, R^\alpha{}_{(\mu} \psi_{|\alpha| \nu)} - 
\frac{4}{3} \, R^\alpha{}_\mu{}^\gamma{}_\nu \, \psi_{\alpha \gamma} +
\frac{1}{6} \, R \, \psi_{\mu \nu} = 0.
\label{CIFE}
\ee
This is the conformal invariant field equation for a symmetric second-rank tensor $\psi_{\mu \nu}$ \cite{CInv4,EO}.

\section{Consistency problems, and a possible solution}
\label{ConSol}

Although conformal invariant, the field equation~(\ref{CIFE}) has consistency problems. In fact, it is not invariant under the transformations
\be
\psi_{\mu \nu} \to \psi_{\mu \nu} - \nabla_\mu \varepsilon_\nu - \nabla_\nu \varepsilon_\mu,
\ee
usually called gauge transformations. As a consequence, even with the non-minimal coupling of the field to the curvature, it is not possible to remove all spurious components of the field, and it turns out to involve more components than necessary to describe a massless field. On the other hand, as discussed in the introduction, a symmetric second-rank tensor is not the only way to represent a spin 2 field: it can also be represented by a 1-form assuming values in the Lie algebra of the translation group
\be
\psi_\nu = \psi^a{}_\nu \, P_a,
\label{S2FieldDef}
\ee
with $P_a = \partial_a$ the translation generators. This second possibility is related to the existence of the tetrad representation of gravity: whereas $\psi_{\mu \nu}$ is conceptually similar to a perturbation of the metric, $\psi^a{}_\nu$ is conceptually similar to a perturbation of the tetrad,
\be
h^a{}_\nu = e^a{}_\nu + \psi^a{}_\nu,
\ee
with $e^a{}_\nu$ a trivial tetrad representing the Minkowski spacetime. Of course, in the same way the metric and the tetrad are equivalent ways of representing the gravitational field, $\psi_{\mu \nu}$ and $\psi^a{}_\nu$ are also equivalent (but different) ways of representing a fundamental spin-2 field. As a matter of fact, it is related to the teleparallel approach to a spin-2 field, according to which gravitation is also represented by a translational-valued gauge potential \cite{livro2}.

Let us then use the Casimir approach to obtain the conformal invariant equation for $\psi^a{}_\nu$. As it is ultimately a (translational-valued) vector field, it is invariant under a conformal transformation:
\be
\bar{\psi}^a{}_\nu = {\psi}^a{}_\nu.
\label{ConTransVec}
\ee
Using the kinematic generators (\ref{lorentzdS}) and (\ref{pi}), as well as the vector matrix generators (\ref{dSM1}) and (\ref{dSM2}), the Casimir operator (\ref{GenCasOp}) for this specific case is found to be
\begin{equation}\label{casimirvec}
(\hat{\mathcal C}_{dS})_\mu{}^\nu = \Omega^{-2} \eta^{\alpha\beta} 
\partial_\alpha \partial_\beta \delta_\mu{}^\nu +
\frac{\Omega^{-1}}{l^2} (x^\nu \partial_\mu -
\eta_{\mu \lambda} \eta^{\nu \rho} x^\lambda \partial_\rho) +
\frac{\Omega^{-1}}{l^2} \delta_\mu{}^\nu -
\frac{1}{2l^4} \eta_{\mu \lambda} x^\lambda x^\nu,
\end{equation} 
with $\Omega$ the conformal factor (\ref{n}). In terms of the Laplace-Beltrami operator acting on $\psi_\nu$, it assumes the form
\begin{equation}
(\hat{\mathcal C}_{dS})_\mu{}^\nu \psi_\nu =
\Box \psi_\mu - R_\mu{}^\nu \, \psi_\nu,
\end{equation}
where $R_\mu{}^\nu$ is the Ricci curvature tensor obtained from (\ref{47}). Substituting the identity
\be
R_\mu{}^\nu \, \psi_\nu = [\nabla^\nu, \nabla_\mu] \psi_\nu,
\label{CurveIdentity2}
\ee
it becomes
\begin{equation}
(\hat{\mathcal C}_{dS})_\mu{}^\nu \psi^a{}_\nu =
\Box \psi^a{}_\mu  - \nabla^\nu \nabla_\mu \psi^a{}_\nu + \nabla_\mu \nabla^\nu \psi^a{}_\nu,
\label{CasiMir}
\end{equation}
where we have re-introduced the algebraic index of $\psi^a{}_\nu$. Using this Casimir operator, the field equation (\ref{ConInvFE2}) with $\sigma = 1$ yields
\begin{equation}
\Box \psi^a{}_\mu - \nabla^\nu \nabla_\mu \psi^a{}_\nu = 0.
\label{S2CIFE2}
\end{equation}
This is the conformal invariant field equation for the translational-valued vector field $\psi^a{}_\nu$. In addition to be conformal invariant, it is also invariant under the gauge transformations
\be
\psi^a{}_\nu \to \psi^a{}_\nu - \partial_\nu \varepsilon^a.
\ee
The gauge invariance, together with the invariance under local Lorentz transformation, render the theory fully consistent.

The field equation (\ref{S2CIFE2}) is written in the class of frames in which no inertial effects are present.\footnote{In the context of teleparallel gravity, where inertia and gravitation are represented by different connections, this class of frames can be easily defined. It is actually the class that reduces to the inertial class of frames in absence of gravitation \cite{livro2}.} Since these effects are represented by a Lorentz connection, which we denote $\Aw^a{}_{b \mu}$, in that specific class of frames such connection vanishes. For this reason, the field equation (\ref{S2CIFE2}) is not manifestly local Lorentz invariant. To rewrite it in a manifestly Lorentz invariant form, it is necessary to perform a local Lorentz transformation
\be
\psi'^a{}_\nu \to \Lambda^a{}_b(x) \, \psi^b{}_\nu.
\ee
In this case, the field equation assumes the form
\be
\daw \psi^a{}_\mu - \nablaw^\nu \nablaw_\mu \psi^a{}_\nu = 0,
\label{LorConfInv}
\ee
where
\be
\nablaw_\mu \psi^a{}_\nu = \nabla_\mu \psi^a{}_\nu + \Aw^a{}_{b \mu} \, \psi^b{}_\nu
\ee
is a covariant derivative that includes a term in the purely inertial connection \cite{spin2}
\be
\Aw^a{}_{b \mu} = \Lambda^a{}_c(x) \, \partial_\mu \Lambda_b{}^c(x).
\label{InerLorConn}
\ee
In this form its local Lorentz invariance becomes manifest. The gauge transformations that leave the field equation~(\ref{LorConfInv}) invariant are now given by
\be
\psi^a{}_\nu \to \psi^a{}_\nu - (\partial_\nu \varepsilon^a +
\Aw^a{}_{b \nu} \, \varepsilon^b).
\ee
Since the inertial Lorentz connection (\ref{InerLorConn}) is invariant under a conformal rescaling of the metric, the local Lorentz invariant field equation (\ref{LorConfInv}) preserves the conformal invariance of the original equation (\ref{S2CIFE2}).

One may wonder why the theory for $\psi_{\mu \nu}$ results inconsistent, whereas the theory for $\psi^a{}_\nu$ turns out to be consistent. The reason is related to the different nature of the indices: whereas a spacetime index has to do with gravitation, the translational algebraic index has to do with inertial effects of the frame only, not with gravitation. As a consequence, the gravitational coupling prescription will be different in each case: whereas the coupling prescription of $\psi_{\mu \nu}$ includes a {\em gravitational connection} term for each index of the field, the coupling prescription of $\psi^a{}_\nu$ includes a {\em gravitational connection} term for the spacetime index, and a {\em purely inertial connection} term for the algebraic (or gauge) index. This difference between the two cases is responsible for rendering the theory, inconsistent in one case, but consistent in the other case.

One should not be surprised with the limitation of the metric-related spin-2 field $\psi_{\mu \nu}$ in describing a fundamental spin-2 field. Remember that, although the gravitational interaction of tensor fields can be dealt with the metric formulation of gravity, the gravitational interaction of spinor fields requires a tetrad formalism \cite{dirac}. The tetrad formulation can then be considered more fundamental than the metric formulation in the sense that it is able to describe the gravitational interaction of both tensor and spinor fields. Analogously, the tetrad-related spin-2 field $\psi^a{}_\nu$ can also be considered more fundamental than $\psi_{\mu \nu}$.

\section{Conclusions}
\label{F}

Minkowski is a homogeneous space, transitive under spacetime translations, whose kinematics is ruled by the Poin\-ca\-r\'e group. The first Casimir operator of the Poincar\'e group yields the d'Alembertian operator in Minkowski spacetime. When equaled to its eigenvalue, it gives the flat spacetime Klein-Gordon equation satisfied by a scalar field. The de Sitter spacetime is also a homogeneous space, but transitive under a combination of translations and proper conformal transformations. The replacement of Minkowski by de Sitter, therefore, naturally introduces the conformal transformations in the spacetime kinematics. Owing to this property, the first Casimir operator of the de Sitter group gives rise to conformal invariant field equations, not only for the scalar field, but for any massless spin-${\sf s}$ field. One has just to use the appropriate representations for each field. Since these equations are covariant and true in de Sitter, they will also be true in any spacetime with non-constant curvature. This procedure constitutes a new, constructive method of obtaining conformal invariant field equations.

Using this approach, we have obtained the conformal invariant equation for a symmetric second-rank tensor $\psi_{\mu \nu}$, given by the field equation~(\ref{CIFE}).
Although conformal invariant, however, that field equation still has the same consistency problems present in its non-conformal invariant version. In fact, it remains not gauge invariant, and consequently involves more independent components than necessary to describe a massless field. The addition of non-minimal coupling terms of the field to the curvature, necessary to implemented the conformal invariance, do not cure the gauge invariance problem \cite{DeserMarc}. On the other hand, a spin-2 field can also be represented by a 1-form assuming values in the Lie algebra of the translation group: $\psi_\nu = \psi^a{}_\nu P_a$. This possibility is related to the existence of the tetrad formulation of gravity: whereas $\psi_{\mu \nu}$ is conceptually similar to a perturbation of the metric, $\psi^a{}_\nu$ is conceptually similar to a perturbation of the tetrad. Of course, in the same way the metric and the tetrad are equivalent ways of representing the gravitational field, $\psi_{\mu \nu}$ and $\psi^a{}_\nu$ are also equivalent (but different) ways of representing a fundamental spin-2 field. Using again the Casimir operator approach, we have obtained the conformal invariant equation for the translational-valued 1-form $\psi^a{}_\nu$, whose manifestly local Lorentz invariant form is given by the field equation~(\ref{LorConfInv}). In addition to be conformal invariant, it is also gauge invariant and consequently fully consistent. These results suggest that a spin-2 field should be interpreted, not as a symmetric second-rank tensor, but as a translational-valued vector field --- a conclusion consistent with the basic principles of teleparallel gravity, a gauge theory for the translation group.

We remark finally that the interpretation of a spin-2 field as a translational-valued vector field finds support in the theory of Lorentz representations. Usually, relying on general relativity, the excitation 2-form of a spin-2 field is assumed to be a curvature-like tensor ${\mathcal R}^{\alpha \beta}{}_{\mu \nu}$. Since this field excitation belongs to the representation
\[
(2, 0) \oplus (0, 2)
\]
of the Lorentz group, it describes spin-2 waves with helicity $\sigma = \pm 2$. On the other hand, relying on the Fierz-Pauli approach to a fundamental spin-2 field \cite{FP} --- or equivalently, on teleparallel gravity --- the excitation 2-form of a spin-2 field can alternatively be assumed to be a Fierz-like tensor ${\mathcal F}_a{}^{\mu \nu}$ \cite{FierzTele}, a field belonging to the representation
\[
(3/2, 1/2) \oplus (1/2, 3/2)
\]
of the Lorentz group \cite{spin2Bis}, which describes spin-2 waves with helicity $\sigma = \pm 1$.\footnote{Gravitational waves with helicity $\sigma = \pm 1$ have already been discussed in Ref.~\cite{vilasi}.} Although not usually considered in the study of spin-2 fields, the field equation governing the dynamics of such waves does not present the consistency problems of the $(2, 0) \oplus (0, 2)$ case.

\section*{Acknowledgments}
The authors would like to thank Yu.\ Obukhov for useful discussions. They would like to thank also FAPESP, CNPq and CAPES for partial financial support. Some of the calculations of the paper were either made or checked with the help of the computer package Cadabra \cite{cada2}.

\end{document}